\documentclass[9pt,twocolumn,twoside]{opticajnl}
\journal{opticajournal}

\setboolean{shortarticle}{true}

\usepackage{lineno}
\usepackage{upgreek}
\usepackage{amsmath}

\newcommand{\minisection}[1]{%
    \par
    \refstepcounter{section}%
    \noindent\textbf{\thesection. #1}%
}

\title{Compressive microendoscopy in the presence of noise and misalignment: the translational correlation advantage of graded-index fibers}

\author[1,2,*]{M. S. Pochechuev}
\author[1,2]{T. I. Rajib}
\author[1,2]{A. Dharmasiri}
\author[1,2]{A. M. Zheltikov}

\affil[1]{Institute for Quantum Science and Engineering, Texas A\&M University, College Station, Texas 77843, USA}
\affil[2]{Department of Physics and Astronomy, Texas A\&M University, College Station, Texas 77843, USA}

\affil[*]{matveymoo@tamu.edu}

\usepackage{xurl}

\makeatletter
\AtBeginDocument{%
  \@ifpackageloaded{hyperref}{\hypersetup{hidelinks}}{}%
}
\makeatother

\begin{abstract}

Multimode fibers (MMFs) offer a minimally invasive platform for deep-tissue and multisite \emph{in vivo} optical sensing. Computational microendoscopy methods extend them into the high-resolution imaging domain, but are quite sensitive to experimental errors. This study investigates the robustness of microendoscopy based on compressive sensing (CS) by comparing step-index (SI) MMFs with graded-index (GRIN) fibers. Mode-expansion-based simulations are used to assess the accuracy of image reconstruction in the presence of fundamental photon noise and lateral fiber shifts. We show that both fiber types exhibit comparable noise sensitivity, but GRIN fibers are significantly more robust to lateral misalignment errors. These results establish GRIN MMFs as an alignment-tolerant platform for robust computational microendoscopy.
\end{abstract}

\setboolean{displaycopyright}{false}

\doi{}
\dates{}
\begin{document}

\maketitle

\noindent\textbf{Introduction.}
Optical fibers provide an ultrathin, minimally invasive platform for deep-tissue optical studies and multisite measurements \cite{kimSimultaneousFastMeasurement2016,turtaev2018highfidelity}. Imaging through multimode fibers, however, requires overcoming the scrambling of light caused by modal interference and usually relies on sophisticated and fragile techniques. On one hand, this can be done by precise control of the transmitted field using wavefront shaping, transmission-matrix measurements, or phase conjugation \cite{revCizmar2012,revPapadopoulos2012,ploschner2015seeing}. On the other hand, computational approaches can recover images from calibrated intensity measurements, shifting part of the complexity from hardware to software \cite{hu2026computational,cao2023controlling}.

Speckle-based CS uses known illumination patterns and a sparsity prior to reconstruct images from fewer measurements than image pixels \cite{candesNearOptimalSignalRecovery2006a,donohoCompressedSensing2006}. Fiber implementations have demonstrated two-dimensional \cite{revAmitonova2018,revAmitonova2020} and three-dimensional imaging \cite{revLyu2024} through SI MMFs. Similarly to wavefront-control methods, CS relies on calibrated, reproducible light propagation and can be very sensitive to experimental changes. Compressive MMF imaging with tolerance to bending has been demonstrated using special optical arrangements \cite{lan2019robust}.
Repeated MMF imaging after fiber reinsertion might require adaptive iterative shift correction \cite{schmidt2022repeated}. Additionally, when the signal is noisy, multiplexed illumination can become disadvantageous compared to standard point-scanning imaging \cite{scottePhotonnoiseSinglepixelCamera2023}. The existing theoretical recovery bounds for signal noise \cite{revCandes2006,needell2013near} and measurement matrix perturbations \cite{herman2010general} establish robustness under stated assumptions, but do not directly quantify the relative performance of different fibers in a specific experimental setting.

Unlike commonly used SI fibers, GRIN fibers have modes with nearly equally spaced groups of propagation constants and can periodically reproduce an input image \cite{snyderOpticalWaveguideTheory1984}. Short pieces of GRIN fiber have been used as imaging probes \cite{satoReflectanceImagingGradedIndex2013,satoImagingCharacteristics882016}. Their ordered modal structure has also been linked to robustness against bending \cite{flaes2018robustness}. We previously demonstrated widefield imaging \emph{in vivo} through implanted telecom GRIN fibers and characterized the length-dependent aberrations that limit this capability \cite{pochechuevImplantableGradedindexFibers2022}. More recently, we investigated CS imaging through these fibers and outlined possible advantages of combining CS reconstruction with standard widefield microscopy \cite{pochechuev2026high}.

In this letter, through numerical simulations, we compare the sensitivity of CS microendoscopy in SI and GRIN fibers to two experimental errors: photon shot noise and lateral input shifts. We demonstrate that both fibers have comparable noise sensitivity at matched photon budgets, but GRIN is substantially more resilient to lateral misalignment. We connect this difference to the translational correlations of the output speckles, which reflect the partial imaging capabilities of GRIN fibers. This result supports the use of GRIN MMFs for robust computational microendoscopy when long-term repeatable operation is required.

\minisection{Imaging model.}
\label{sec:ImModel}
We simulate CS imaging using the fiber model developed in our previous work \cite{pochechuev2026high}, which qualitatively reproduced the GRIN aberrations observed experimentally \cite{pochechuevImplantableGradedindexFibers2022}. Here we add an SI fiber with the same core diameter $D=50~\upmu\mathrm{m}$ and numerical aperture $\mathrm{NA}=0.22$. We calculate propagation over lengths up to $8~\mathrm{mm}$ using orthonormal scalar guided modes $E_j$ and propagation constants $\beta_j$. More details are given in Supplement 1, Sections 1 and 2.

\begin{figure}[!t]

\centering
 \includegraphics[width=1\linewidth]{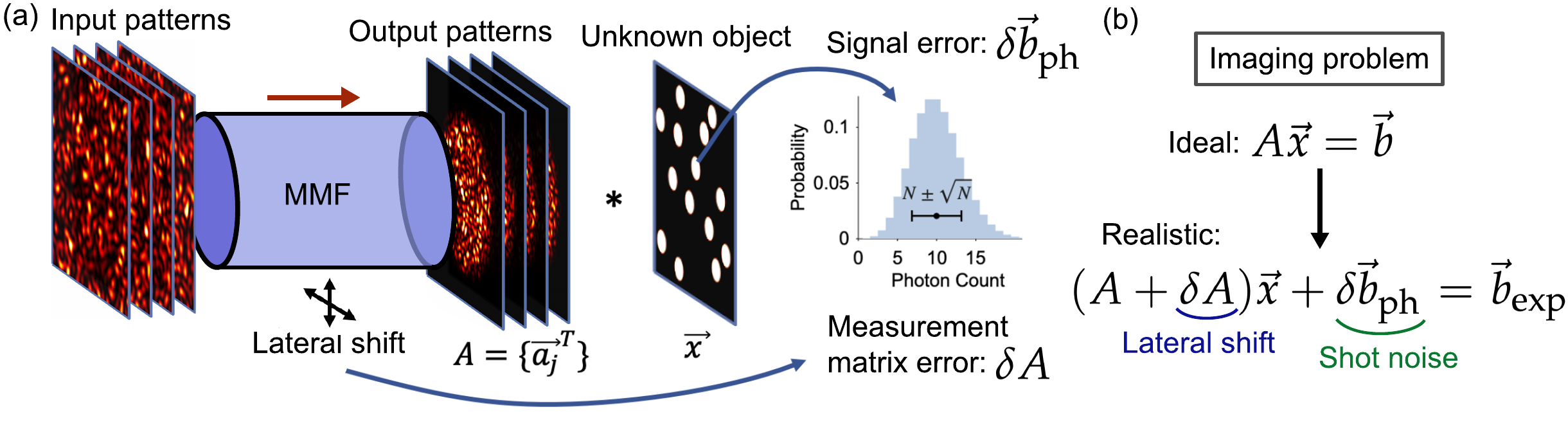}
\caption{Speckle-based CS microendoscopy and the two types of error considered. (a) Input speckle patterns go through the MMF, resulting in output patterns $\{\vec{a}_i^{\,T}\}$ that form the rows of the measurement matrix $A$. A relative lateral shift leads to the matrix error $\delta A$, while shot noise leads to a signal error $\delta\vec{b}_{\mathrm{ph}}$. Inset: example Poisson distribution of photon counts with average $10$ photons/measurement. (b) Modification of the ideal imaging problem due to lateral shifts and shot noise.}
\label{fig:setup}
\end{figure}

The idea of CS microendoscopy is shown in Fig.~\ref{fig:setup}. A series of illumination speckle patterns is sent into the MMF, where the patterns evolve during propagation. The output speckle patterns $\{\vec{a}_i\}$ illuminate the object $\vec{x}$, generating the fluorescence response $\vec{b}$. Here we assume discretization of the two-dimensional spatial distributions and reordering into one-dimensional vectors. The imaging problem can then be summarized as $A\vec{x}=\vec{b}$, where the measurement matrix $A$ has rows $\vec{a}_i^{\,T}$ and size $M\times N$: $M$ is the number of illumination patterns and $N$ the number of image pixels. The experimentally measured signal $\vec{b}_{\mathrm{exp}}$ can differ from the exact expected signal $\vec{b}$ in the presence of noise $\delta\vec{b}_{\mathrm{ph}}$ and a measurement-matrix perturbation $\delta A$: $\vec{b}_{\mathrm{exp}}=(A+\delta A)\vec{x}+\delta\vec{b}_{\mathrm{ph}}$. The CS approach seeks a reconstruction with $M<N$ by solving an optimization problem:
\begin{equation}
    \hat{\vec x}=\underset{\vec z}{\operatorname{arg\,min}}\,\mathrm{TV}(\vec z)
    \quad\text{subject to}\quad
    \|A\vec z-\vec b_{\mathrm{exp}}\|_2\leq\epsilon,
    \label{eq:CS_problem}
\end{equation}
where $\hat{\vec x}$ is the reconstructed object, $\mathrm{TV}$ denotes total variation \cite{rudin1992nonlinear,needell2013near}, and $\epsilon$ is the allowed data mismatch. In this work, we numerically investigate the effects of signal noise and shift-induced matrix errors on the quality of the CS solution. In all cases, $\epsilon$ is chosen to match the combined residual, $\epsilon=\|\vec b_{\mathrm{exp}}-A\vec x\|_2$.
To estimate the effects of these errors separately and simultaneously, we numerically simulate the CS experiment by generating the object $\vec{x}$, illumination patterns forming the measurement matrix $A$, and experimental signal $\vec b_{\mathrm{exp}}$. We use the L1-magic solver \cite{L1Magic} and average repeated simulations to determine practical limits of CS microendoscopy in SI and GRIN fibers. Further simulation details are given in Supplement 1, Section 4.

\minisection{Translational Correlations in MMFs.}
\label{sec:TrCor}
In CS-based microendoscopy, the measurement matrix $A$ is measured during calibration and later used in experiments. In many relevant settings, such as long-term brain imaging, the fiber needs to be reinstalled in the optical system many times. Reinstallation can cause unknown shifts of the fiber with respect to the input speckle fields, leading to a perturbation $\delta A$ of the measurement matrix. To study this perturbation, we consider the effects of a shift on the propagation of speckle patterns.

Let $\mathcal{P}_L$ denote propagation over distance $L$, and $\mathcal{T}_{\Delta \mathbf{r}}$ denote transverse translation by ${\Delta\mathbf{r}=(\Delta x,\Delta y)}$ such that $E_{\mathrm{out}}(\mathbf{r})=\mathcal{P}_L E_{\mathrm{in}}(\mathbf{r})$ and $\mathcal{T}_{\Delta \mathbf{r}}E(\mathbf{r})=E(\mathbf{r}-\Delta\mathbf{r})$. For free-space propagation or shift-invariant imaging with unit magnification, propagation and spatial shifts commute ($[\mathcal{P}_L,\mathcal{T}_{\Delta\mathbf{r}}]=0$). The noiseless measurements then satisfy $(A+\delta A)\vec x=A\mathcal{T}_{-\Delta\mathbf r}\vec x=\vec b_{\mathrm{exp}}$, provided that cropping and boundary effects are negligible. Reconstruction with the original $A$ can therefore recover a shifted object rather than a distorted one. Thus, the effect of a matrix perturbation depends on its structure, rather than its magnitude alone.

In a multimode waveguide, shifting the input redistributes energy among the guided modes, which generally have different propagation constants. Their mismatch leads to decorrelation of the input field over distances $L_{\mathrm{dec}}\approx2\pi/\Delta\beta$, where $\Delta\beta$ is the effective propagation-constant difference of the excited modes (see Supplement 1, Section 2). Translational correlations, however, depend on the commutation between propagation and translation operators. To quantify them, we compare two fields at the output: the first generated by a physically shifted input ($E_1=\mathcal{P}_L\mathcal{T}_{\Delta\mathbf r}E_{\mathrm{in}}$) and the second mathematically shifted after propagation ($E_2=\mathcal{T}_{\Delta\mathbf r}\mathcal{P}_LE_{\mathrm{in}}$). We assume all input fields to be normalized to unit power: $\int E_{\mathrm{in}}^*(\mathbf r)E_{\mathrm{in}}(\mathbf r)\,d^2\mathbf r=1$. One way to compare the fields is by evaluating their difference: $E_1-E_2=[\mathcal{P}_L,\mathcal{T}_{\Delta\mathbf r}]E_{\mathrm{in}}$. Another way is to consider their field overlap $M_E(\Delta\mathbf r,L)=\int E_2^*(\mathbf r)E_1(\mathbf r)\,d^2\mathbf r$. For CS reconstruction, we are specifically interested in speckles as illumination patterns. We generate speckles as $E_{\mathrm{in}}=\sum_jc_jE_j$ with equal magnitudes and independent random phases, so that $c_j=e^{i\phi_j}/\sqrt{N_{\mathrm{modes}}}$. We quantify the translational correlations by the ensemble-averaged overlap integral:
\begin{equation}
    \langle M_E(\Delta \mathbf{r},L)\rangle=\frac{1}{N_{\mathrm{modes}}}
    \sum_{m,n}|T_{mn}(\Delta\mathbf r)|^2e^{i(\beta_m - \beta_n)L}.
\label{eq:AvME}
\end{equation}
Here we used $\langle c_k^*c_l\rangle=\delta_{kl}/N_{\mathrm{modes}}$ and the matrix elements of the shift operator $T_{mn}=\int E_m^*(\mathbf r)E_n(\mathbf r-\Delta\mathbf r)\,d^2\mathbf r$.
Note that $\langle M_E(\Delta \mathbf{r},L)\rangle$ is strictly real. For small shifts $\Delta\mathbf r$, the losses are negligible and we find the following relation to the commutator: $\langle\int|[\mathcal P_L,\mathcal T_{\Delta\mathbf r}]E_{\mathrm{in}}|^2\,d^2\mathbf r\rangle\approx2-2\langle M_E\rangle$. When the commutator is zero, the overlap equals one, $\langle M_E\rangle=1$, whereas for completely uncorrelated fields $\langle M_E\rangle=0$ and the squared field difference is approximately two (for details see Supplement 1, Section 3).

Since fluorescence excitation depends on intensities rather than fields, we also calculate $M_I=\rho(|E_1|^2,|E_2|^2)$, where $\rho$ is the spatial Pearson correlation, and average it over speckle realizations. While $M_E$ allows for simpler analytical treatment, $M_I$ helps directly test whether the shifted illumination retains its intensity structure. Our results show that, for the speckle fields considered, $M_E$ and $M_I$ qualitatively describe the same behavior.

\begin{figure*}[t]
\centering
 \includegraphics[width=0.86\textwidth]{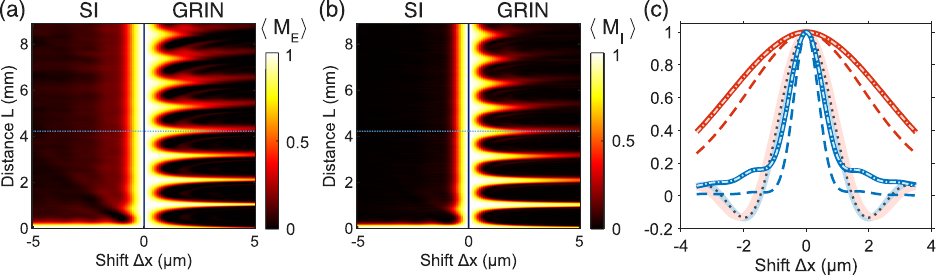}
\caption{Translational correlations in SI and GRIN fibers. Field-based (a) and intensity-based (b) ensemble-averaged translational correlations $\langle M_E(\Delta\mathbf r,L)\rangle$ and $\langle M_I(\Delta\mathbf r,L)\rangle$ displayed as heatmaps for $\Delta\mathbf r=(\Delta x,0)$. (c) Profiles of $\langle M_E\rangle$ (solid) and $\langle M_I\rangle$ (dashed) taken at $L=4.23~\mathrm{mm}$ calculated via averaging over an ensemble of speckles. Dash-dotted lines show the analytical prediction for $\langle M_E\rangle$. Ensemble-averaged field autocorrelation functions for speckles in SI (light blue), GRIN (light red) and free space (dotted) are shown for reference.}
\label{fig:fibers}
\end{figure*}

Figure~\ref{fig:fibers} compares the ensemble-averaged field and intensity translational correlations, $\langle M_E(\Delta\mathbf r,L)\rangle$ and $\langle M_I(\Delta\mathbf r,L)\rangle$, in SI and GRIN fibers. The heatmaps in Figs.~\ref{fig:fibers}(a) and \ref{fig:fibers}(b) demonstrate the key difference between the two types of fiber. For the nonzero shifts shown, the SI correlations decay after propagation over just $L\approx500~\upmu\mathrm m$. Conversely, the GRIN fiber demonstrates periodic revivals of translational correlations, leading to an increased shift range. This effect is directly related to the periodic refocusing property of GRIN fibers near integer pitch lengths, $L=mL_{\mathrm{pitch}}$, where $m$ is an integer and $L_{\mathrm{pitch}}=2\pi/g\approx\pi n_0D/\mathrm{NA}\approx1.06~\mathrm{mm}$ in our model; $g\approx2\mathrm{NA}/(n_0D)$ is the parabolic-profile gradient and $n_0$ the on-axis refractive index. Figure~\ref{fig:fibers}(c) shows heatmap cross-sections at $L=4L_{\mathrm{pitch}}\approx4.23~\mathrm{mm}$, demonstrating a wider correlation range in GRIN. It also demonstrates close agreement between the ensemble-average estimation and the analytical prediction from Eq.~(\ref{eq:AvME}). Additionally, we plot the ensemble-averaged speckle autocorrelation for GRIN, SI and free-space speckles of the same numerical aperture. The autocorrelation $M_E^{\mathrm{auto}}(\Delta\mathbf r)=\int E_{\mathrm{in}}^*(\mathbf r)\mathcal T_{\Delta\mathbf r}E_{\mathrm{in}}(\mathbf r)\,d^2\mathbf r$ characterizes the intrinsic spatial correlations of the speckle field, and its characteristic width is of the order of the diffraction-limited resolution \cite{goodman}.
Interestingly, while the phase factors $e^{i(\beta_m-\beta_n)L}$ drive the decorrelation, the higher weights $|T_{mn}(\Delta\mathbf r)|^2$ in Eq.~(\ref{eq:AvME}) favor smaller propagation-constant differences, leading to a substantially longer translational-correlation decay length than the input--output scale $L_{\mathrm{dec}}\approx2\pi/\Delta\beta$ (for a comparison between different correlations, see Supplement 1, Section 3).

Thus, translational correlations are preserved for longer propagation distances than the correlation between the output and input fields. This suggests that CS can still benefit from the modal structure of GRIN fibers when standard microscopy no longer works due to strong aberrations. While this effect is observed in both fibers, GRIN preserves a wider correlation range over the millimeter lengths relevant to deep-tissue endoscopy. We discuss the benefits of this extended correlation range in Section~\ref{sec:Results}.

\minisection{Noise Sensitivity.}
\label{sec:Noise}
For a static, noiseless system, CS theory predicts nearly perfect recovery with a number of measurements $M$ scaling with sparsity under suitable assumptions on the sensing matrix and object \cite{candesRobustUncertaintyPrinciples2006,needell2013near}. Importantly, in practical imaging systems, these assumptions often do not hold, even though CS successfully reconstructs objects \cite{bastounisAbsenceUniformRecovery2017}. In our simulations, we therefore chose $M$ empirically for the aligned, noise-free system to achieve high-quality reconstruction with $Q\simeq0.995$ for both fibers, then held it fixed while varying other parameters to study the robustness of the method (Supplement 1, Section 4).

In a real experiment, any measurement is corrupted by technical and fundamental errors. Here we assume that the detection device is ideal and there are no other sources of error except fundamental photon shot noise. In this limit, at zero shift, $\vec b=A\vec x$ is the expected signal, while the measured signal $\vec b_{\mathrm{exp}}$ is obtained from Poisson-distributed photon counts \cite{raginsky2010compressed}. The signal error is then $\delta\vec b_{\mathrm{ph}}=\vec b_{\mathrm{exp}}-\vec b$. Conveniently, in this model we can describe the noise level by a single parameter: the total expected number of detected photons $N_{\mathrm{ph}}$. Higher $N_{\mathrm{ph}}$ corresponds to lower relative signal noise (Supplement 1, Section 4).

\begin{figure*}[ht]
\centering
 \includegraphics[width=0.92\textwidth]{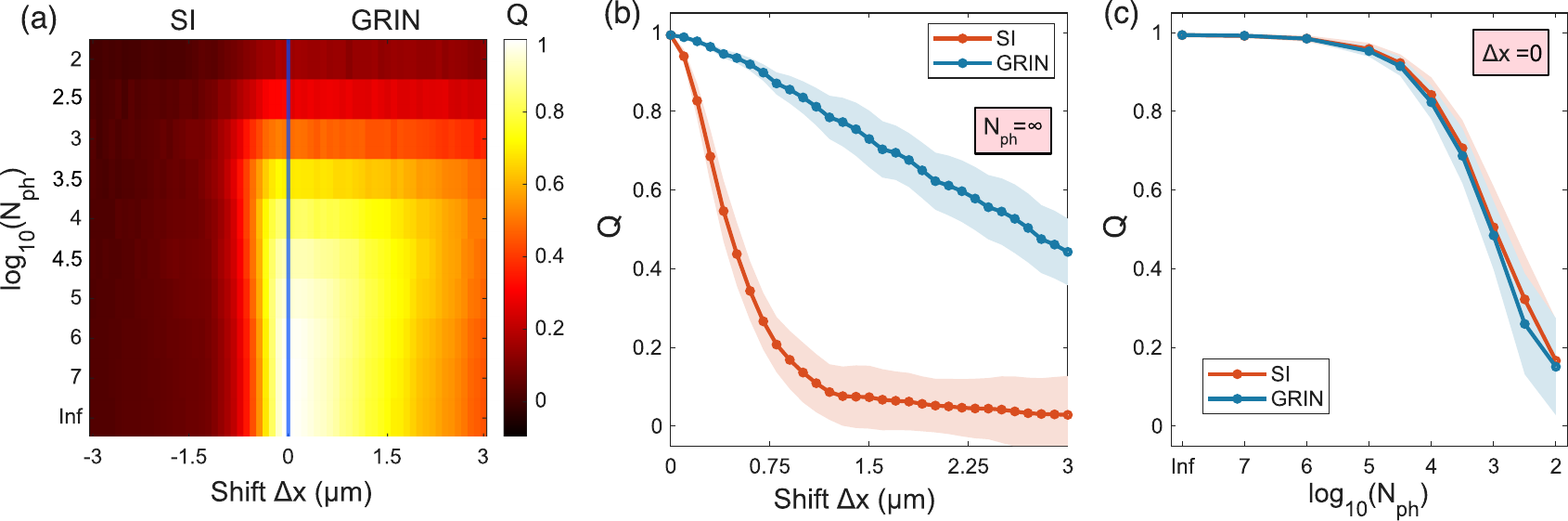}
\caption{CS reconstruction quality under photon noise, specified by the expected total detected photon count $N_{\mathrm{ph}}$, and lateral input shifts $\Delta x$ in SI and GRIN fibers. (a) Heatmap of mean Pearson correlation $Q(\Delta x,N_{\mathrm{ph}})$ between the reconstructed and ground-truth images. (b) Noiseless profiles $Q(\Delta x,\infty)$ demonstrate greater shift tolerance in GRIN than SI. (c) Zero-shift profiles $Q(0,N_{\mathrm{ph}})$ show practically identical photon-noise sensitivity in the two fiber types. Shaded bands correspond to standard deviation.}
\label{fig:QvsShiftvsNph}
\end{figure*}

In addition to signal noise, the error associated with the shift-induced measurement-matrix perturbation $\delta A$ can be viewed as an object-dependent error vector $\delta\vec b_A=\delta A\vec x$ and therefore the same bounds can be used to guarantee reconstruction. However, as discussed in the previous section, in many situations $\delta A$ retains structure due to translational correlations, which can be beneficial for robust CS microendoscopy. We study the effects of both errors separately and simultaneously by estimating CS reconstruction quality via averaging over many speckle realizations and Poisson noise instances. The main metric we use to quantify reconstruction quality is the Pearson correlation $Q=\rho(\vec x,\hat{\vec x})$ (Eq.~(S11)), which measures similarity of image structure and is insensitive to positive overall intensity scaling. $Q=1$ indicates perfect correlation and $Q=0$ no linear correlation.

\minisection{Results and Discussion.}
\label{sec:Results}
To assess the effects of lateral shifts and photon noise on CS reconstruction in SI and GRIN fibers, we ran a series of simulations. The results are summarized in Fig.~\ref{fig:QvsShiftvsNph}.
The number of measurements was fixed at a sufficiently high value as explained in Supplement 1, Section 4. The heatmap (Fig.~\ref{fig:QvsShiftvsNph}(a)) gives a full comparison of SI and GRIN performance for shifts up to $\Delta x=3~\upmu\mathrm m$ and $N_{\mathrm{ph}}$ from $10^2$ to $\infty$ (no noise). As shown in Section~\ref{sec:TrCor}, for the SI fiber the typical correlation-preserving shifts are comparable to the diffraction-limited spatial scale. The CS reconstruction quality $Q$ shows similar behavior, falling to its minimum at about $\Delta x=1.5~\upmu\mathrm m$ (Fig.~\ref{fig:QvsShiftvsNph}(b)), while the GRIN fiber has a range at least six times wider, measured at $Q=0.5$.

Shift correlations considered in this work are a form of optical memory effect \cite{judkewitz2015translation,osnabrugge2017generalized}. Memory effects in MMFs have also been exploited for imaging and transmission-matrix estimation \cite{li2021memory,li2021compressive}. Using a GRIN fiber substantially increases the translational memory range, relaxing the requirements on reinstallation accuracy compared to SI fiber. Unknown lateral shifts after MMF reinsertion can be compensated by adaptive optics \cite{schmidt2022repeated}. This requires metrics that help distinguish image from noise without requiring knowledge of the ground truth to drive the iterative compensation process. Visual inspection shows that the image is lost at a shift of about $0.5~\upmu\mathrm m$ for SI, while GRIN fiber preserves recognizable details up to $3~\upmu\mathrm m$ in the examples considered (Supplement 1, Fig.~S5, upper part). Moreover, the noisy images tend to have a substantial number of physically meaningless pixels with negative intensity. The intensity-weighted fraction of negative pixels distinguishes recognizable noisy images in GRIN from noise-induced artifacts in SI (Supplement 1, Fig.~S5, lower part). This simple metric could therefore guide manual or automatic shift compensation.

\minisection{Conclusion.}
\label{sec:Conclusion}
CS with speckle illumination enables high-resolution microendoscopy through ultrathin optical fibers. Theoretical recovery bounds provide useful context, but do not directly determine practical tolerance to experimental errors in a particular fiber system. Our results demonstrate that while SI and GRIN fibers exhibit comparable sensitivity to photon shot noise, there is a substantial difference in tolerance to lateral input shifts. The structured modes of GRIN fibers preserve a wider translational-correlation range even when aberrations limit direct imaging. This advantage supports their use in computational microendoscopy where repeatable input alignment is difficult.

\smallskip
\noindent\textbf{Funding.} Welch Foundation (A-1801-20210327); Office of Science (DE-SC0024882).

\smallskip
\noindent\textbf{Acknowledgment.} This work was supported by the Welch Foundation (A-1801-20210327) and the DOE Office of Science through the DOE Inertial Fusion Energy Science and Technology Accelerator Research (IFE-STAR) program (grant \#DE-SC0024882).

\smallskip
\noindent\textbf{Disclosures.} The authors declare no conflicts of interest.
\smallskip

\noindent\textbf{Data availability.} The data underlying the results presented in this paper are not publicly available but may be obtained from the authors upon reasonable request.

\smallskip
\noindent See Supplement 1 for supporting content.

\bibliography{paper_2_curated_upd3}
\bibliographyfullrefs{paper_2_curated_upd3}
\end{document}

\begin{center}

    {\large \textbf{Supplementary Material}}\\[1em]

    {\Large \textbf{Compressive microendoscopy in the presence of noise and misalignment: the translational correlation advantage of graded-index fibers}}\\[1.0em]

    {\large
    M. S. Pochechuev,$^{1,2,*}$
    T. I. Rajib,$^{1,2}$ A. Dharmasiri,$^{1,2}$ and
    A. M. Zheltikov$^{1,2}$
    }\\[1em]

    {\small \textit{
    $^1$Institute for Quantum Science and Engineering, Texas A\&M University, College Station, Texas 77843, USA\\
    $^2$Department of Physics and Astronomy, Texas A\&M University, College Station, Texas 77843, USA
    }}\\

    {\small $^*$matveymoo@tamu.edu}

\end{center}

\section{Differences between SI and GRIN fibers}
\label{sec:fiber_models}

We considered SI and GRIN fibers with the same core diameter $D=2a=50~\upmu\mathrm{m}$, numerical aperture $\mathrm{NA}=0.22$, and maximum refractive index $n_0=1.48$. The laser wavelength was set to $\lambda=532~\mathrm{nm}$. The SI fiber has the core refractive index $n_{\mathrm{core}}=n_0$ at $r<a$ and the cladding refractive index $n_{\mathrm{clad}}=\sqrt{n_0^2-\mathrm{NA}^2}$ at $r\geq a$. The ideal GRIN fiber differs inside the core, which has a parabolic squared refractive-index profile $n_{\mathrm{parab}}^2(r)=n_0^2(1-g^2r^2)$, where $g=2\mathrm{NA}/(n_0D)$ is the gradient constant.
We calculated fiber modes in the scalar paraxial approximation, which is valid at low numerical apertures and short propagation distances. The SI and ideal parabolic GRIN fibers have different mode profiles $E_j(x,y)$, described by Bessel and Laguerre--Gaussian functions, respectively \cite{snyderOpticalWaveguideTheory1984}. The main difference between the two for our work is the distribution of their propagation constants $\beta_j$. The SI modes have propagation constants determined by the core--cladding boundary conditions. They can be found as solutions of a transcendental equation, which leads to a nearly uniform distribution over the whole supported range $kn_{\mathrm{clad}}\leq\beta\leq kn_{\mathrm{core}}$, where $k=2\pi/\lambda$ (see \ref{sec:propagation}). The GRIN propagation constants form mode groups with $\beta_j^{\mathrm{parab}}\simeq kn_0-gq_j$, where $q_j$ is the integer principal mode-group index. To model a realistic GRIN fiber, we introduced refractive-index deviations from the parabolic profile and found the modes and propagation constants using perturbation theory. Throughout the main text, GRIN refers to the perturbed graded-index fiber used in our previous model \cite{pochechuev2026high}.

\begin{figure}[!b]
\centering
 \includegraphics[width=0.745\linewidth]{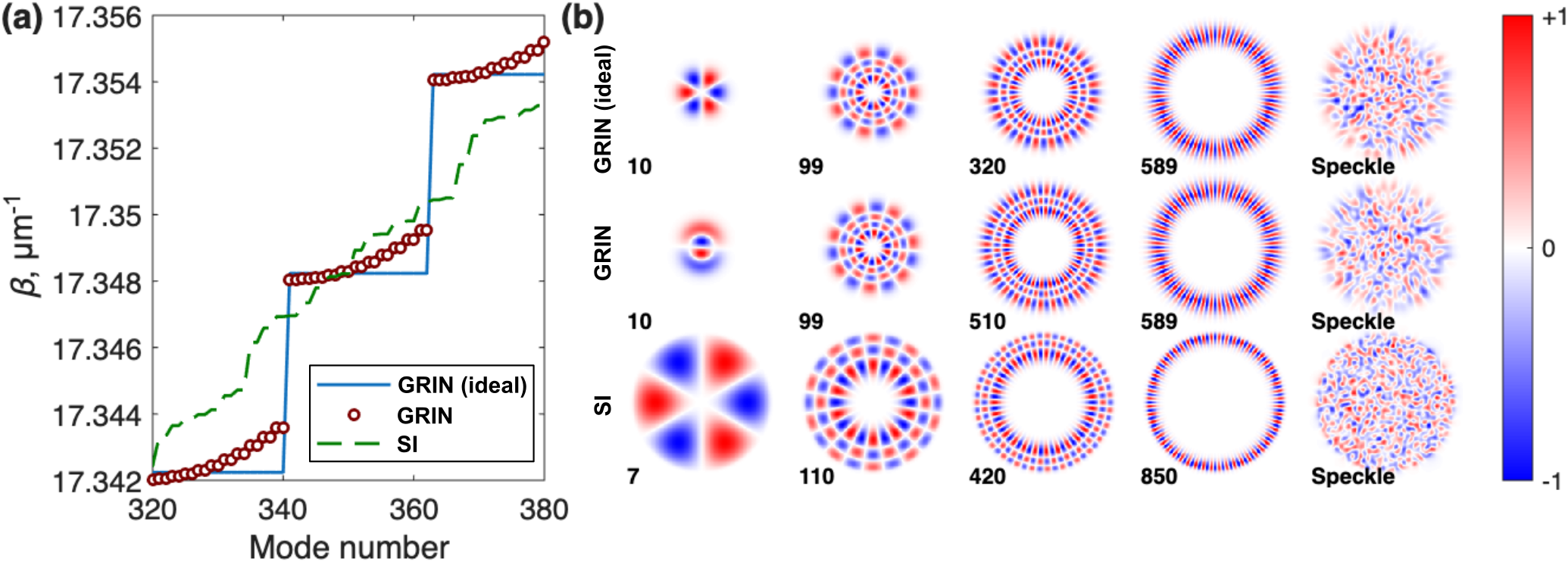}
\caption{Comparison of the modal properties of the three simulated fibers: GRIN (ideal), with a parabolic squared refractive-index profile (top row); GRIN, with an added perturbation; and SI. (a) The ladder-like (GRIN) and nearly uniform (SI) distributions of propagation constants $\beta$. (b) Representative distributions of the real part of the fiber modes. The last column shows a speckle field calculated as a sum of all modes with random phases.}
\label{fig:fibers}
\end{figure}

Figure~\ref{fig:fibers}(a) shows the ladder-like distribution of propagation constants for the ideal GRIN fiber, their deviations for the perturbed GRIN, and the nearly uniform distribution for the SI fiber. Note that in our numerical simulations we indexed the modes by sorting $\beta$ in increasing order, whereas $q_j$ labels a mode group rather than the position in this sorted list. Figure~\ref{fig:fibers}(b) qualitatively compares the mode distributions and representative speckles. In the large-mode-number limit, the SI fiber supports approximately twice as many modes as the GRIN fiber (1061 vs 593 in our model). The perturbation mainly affects the propagation constants, and the differences between representative mode patterns of ideal and perturbed GRIN fibers mainly stem from a different choice of basis within degenerate mode groups. The last column demonstrates speckles constructed as superpositions of modes taken with equal amplitude but random phases. Qualitatively, GRIN and SI speckles are very similar, with the most noticeable differences near the core boundary, where GRIN has a lower effective numerical aperture.

\section{Light propagation in optical fibers}
\label{sec:propagation}
\begin{figure}[!b]
\centering
 \includegraphics[width=0.89\linewidth]{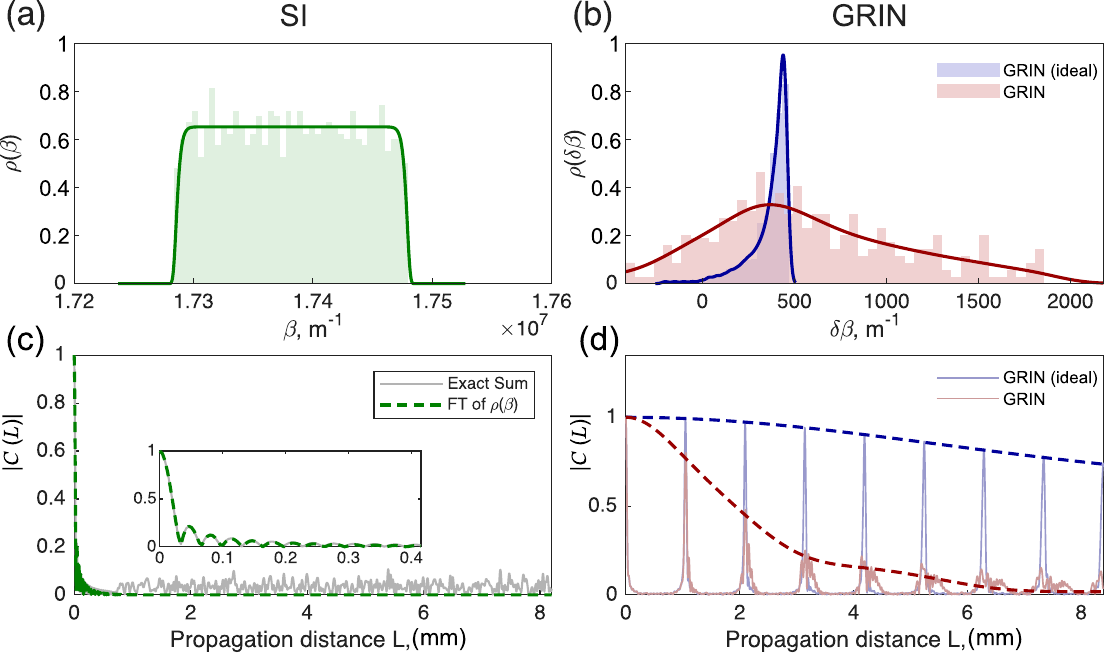}
\caption{Mode-density distributions and corresponding input--output field correlations $C(L)$. (a,b) Normalized densities $\rho(\beta)$ for SI and $\rho(\delta\beta)$ for the ideal and perturbed GRIN fibers. Bar charts show the modal histograms, and solid curves show their smooth approximations using either a super-Gaussian (a) or a continuous kernel density estimator (b). (c) $|C(L)|$ for SI (solid gray) and the magnitude of the Fourier transform of its smooth mode density (dashed green). (d) $|C(L)|$ for GRIN (solid lines) and the magnitude of the Fourier transform of $\rho(\delta\beta)$ that approximates it at $L=mL_{\mathrm{pitch}}$.}
\label{fig:phases}
\end{figure}

To accurately predict the change in the measurement matrix caused by lateral shifts, we model light propagation through a waveguide supporting a set of orthonormal scalar guided modes $\{E_j(\mathbf r)\}$ with corresponding propagation constants $\{\beta_j\}$, where $j$ runs from $1$ to $N_{\mathrm{modes}}$. Propagation over a distance $L$ is given by the linear operator $\mathcal P_L$:
\begin{equation}
\mathcal P_LE_{\mathrm{in}}(\mathbf r)
=\sum_{j=1}^{N_{\mathrm{modes}}}c_jE_j(\mathbf r)e^{i\beta_jL},
\qquad c_j=\int E_j^*(\mathbf r)E_{\mathrm{in}}(\mathbf r)\,d^2\mathbf r,
\qquad \int E_m^*E_n\,d^2\mathbf r=\delta_{mn}.
\label{Eq:Propag}
\end{equation}

To investigate how the intensity distribution changes during propagation through optical fibers, it is easier to first consider the field evolution. For this, we define the normalized input--output field correlation, $C(L)$, for an initially fully guided normalized field $E_{\mathrm{in}}(\mathbf r)$:
\begin{equation}
C(L)=\int E_{\mathrm{in}}^*(\mathbf r)\mathcal P_LE_{\mathrm{in}}(\mathbf r)\,d^2\mathbf r.
\label{eq:input_output}
\end{equation}
The magnitude $|C(L)|$ equals one when the output field reproduces the input up to a global phase and approaches zero for fields with negligible overlap. The evolution of this correlation strongly depends on the modal composition of the input. For an individual fiber mode, $|C(L)|$ remains one. However, for a field comprising multiple modes, $C(L)$ varies due to modal dephasing driven by the differences in propagation constants.

In this study, we are primarily interested in speckle fields, which are modeled as an equal-amplitude superposition of guided modes with independent, uniformly distributed phases: $E_{\mathrm{sp}}=\sum_jc_jE_j$, with $c_j=e^{i\phi_j}/\sqrt{N_{\mathrm{modes}}}$. For such a field, the input--output field correlation simplifies to the coherent sum of modal phasors:
\begin{equation}
C(L)=\frac{1}{N_{\mathrm{modes}}}\sum_j e^{i\beta_jL}.
\label{eq:CL}
\end{equation}
The absolute value $|C(L)|$ is calculated numerically for the ideal and perturbed GRIN fibers, as well as the SI fiber (solid lines in Fig.~\ref{fig:phases}(c,d)). The behavior of $C(L)$ can also be analyzed analytically by considering a large number of modes and approximating the sum with a Fourier integral, $C(L)\simeq\int\rho(\beta)e^{i\beta L}\,d\beta$, where $\rho(\beta)$ is the continuous normalized mode density.
The SI mode density is nearly uniform in the range from $kn_{\mathrm{clad}}$ to $kn_{\mathrm{core}}$. It can be approximated by a high-order super-Gaussian (solid curve in Fig.~\ref{fig:phases}(a)). The Fourier transform of this approximation shows close agreement with the modal calculation at short distances (Fig.~\ref{fig:phases}(c)), exhibiting a sinc-like decay with the first zero near
\begin{equation}
L_0\simeq\frac{2\pi}{\beta_{\max}-\beta_{\min}}
\simeq\frac{2\lambda n_0}{\mathrm{NA}^2}
\simeq33~\upmu\mathrm m.
\label{eq:SI_scale}
\end{equation}
As $L$ increases, the modal phases spread around the complex unit circle and $C(L)$ for SI decreases toward zero. The remaining fluctuations reflect the finite number of modal contributions.
{Their typical magnitude can be estimated by treating the dephased contributions as a random walk, giving an RMS value of order $\tilde{N}_{\mathrm{modes}}^{-1/2}$, where $\tilde{N}_{modes}$ is the number of nondegenerate modes.}

In contrast, the GRIN propagation constants are nearly equidistant and rephase at distances $L=mL_{\mathrm{pitch}}$, where $m$ is an integer and $L_{\mathrm{pitch}}=2\pi/g=\pi n_0D/\mathrm{NA}\simeq1.06~\mathrm{mm}$. To describe decorrelation, it is useful to consider deviations from the ideal ladder, $\delta\beta_j=\beta_j-(kn_0-gq_j)$. At integer pitch lengths, the input--output overlap is
\begin{equation}
C(mL_{\mathrm{pitch}})
=\frac{1}{N_{\mathrm{modes}}}
\sum_j e^{i\delta\beta_j(mL_{\mathrm{pitch}})}.
\label{eq:GRIN_residual}
\end{equation}
Similar to the SI analysis, the deviation distributions $\rho(\delta\beta)$ and their smooth approximations are shown in Fig.~\ref{fig:phases}(b). Their Fourier transforms approximately predict the value of $|C(L)|$ at $L=mL_{\mathrm{pitch}}$. As the intermodal phase mismatch increases, $|C(L)|$ drops. From an imaging perspective, this corresponds to the accumulation of aberrations. For the considered perturbation, $|C(L)|$ drops significantly after just a few millimeters.

\clearpage
\section{Translational correlations during propagation}
\label{sec:translation}

To quantify translational correlations, we compare the output of a shifted input with a shifted copy of the original output, as in the main text. Let $\mathcal T_{\Delta\mathbf r}E(\mathbf r)=E(\mathbf r-\Delta\mathbf r)$ and $T_{mn}=\int E_m^*(\mathbf r)E_n(\mathbf r-\Delta\mathbf r)\,d^2\mathbf r$. For an input speckle $E_{\mathrm{sp}}=\sum_nc_nE_n$, the two fields are
\begin{equation}
\begin{aligned}
E_1&=\mathcal P_L\mathcal T_{\Delta\mathbf r}E_{\mathrm{sp}}
=\sum_{m,n}E_mT_{mn}c_ne^{i\beta_mL},\\
E_2&=\mathcal T_{\Delta\mathbf r}\mathcal P_LE_{\mathrm{sp}}
=\sum_n(\mathcal T_{\Delta\mathbf r}E_n)c_ne^{i\beta_nL}.
\end{aligned}
\label{eq:shifted_fields}
\end{equation}
To assess translational correlations, we consider the field overlap $M_E(\Delta\mathbf r,L)=\int E_2^*E_1\,d^2\mathbf r$. Substitution of Eq.~\eqref{eq:shifted_fields} gives
\begin{equation}
M_E=\sum_{m,k,n}c_k^*c_nT_{mk}^*T_{mn}e^{i(\beta_m-\beta_k)L}.
\label{eq:overlap_before_average}
\end{equation}
Averaging over the ensemble of independent speckles gives
\begin{equation}
\begin{aligned}
\langle M_E(\Delta\mathbf r,L)\rangle
&=\frac{1}{N_{\mathrm{modes}}}
  \sum_m\sum_n |T_{mn}(\Delta\mathbf r)|^2
  e^{i\Delta\beta_{mn}L}\\
&=\frac{1}{N_{\mathrm{modes}}}\sum_n|T_{nn}|^2
  +\frac{2}{N_{\mathrm{modes}}}
  \sum_{m>n}|T_{mn}|^2\cos(\Delta\beta_{mn}L).
\end{aligned}
\label{eq:overlap_after_average}
\end{equation}
Here we used $\langle c_k^*c_n\rangle=\delta_{kn}/N_{\mathrm{modes}}$, $\Delta\beta_{mn}=\beta_m-\beta_n$, $|T_{mn}|^2=|T_{nm}|^2$, and separated the diagonal and off-diagonal contributions.

For small $\Delta\mathbf r$, we can assume that shift-induced boundary and guided-power losses are negligible, so that
$\int |E_1(\mathbf r)|^2\,d^2\mathbf r
\approx\int |E_2(\mathbf r)|^2\,d^2\mathbf r
\approx\int |E_{\mathrm{sp}}(\mathbf r)|^2\,d^2\mathbf r=1$.
Then $\langle M_E(\Delta\mathbf r,L)\rangle$ is related to the commutator $[\mathcal P_L,\mathcal T_{\Delta\mathbf r}]$:
\begin{equation}
\begin{aligned}
\langle\int|[\mathcal P_L,\mathcal T_{\Delta\mathbf r}]E_{\mathrm{sp}}|^2\,d^2\mathbf r\rangle
\approx2-2\langle M_E\rangle.
\end{aligned}
\label{eq:commutator_relation}
\end{equation}

Equation~\eqref{eq:overlap_after_average} shows that $\langle M_E(\Delta\mathbf r,L)\rangle$ is real, and its decay with increasing $L$ is associated with intermodal dephasing through the phase terms $e^{i\Delta\beta_{mn}L}$, similar to Eq.~\eqref{eq:CL}. High translational correlation can be achieved at small $\Delta\mathbf r$, where the matrix $T_{mn}$ stays nearly diagonal ($\sum_n|T_{nn}|^2\approx N_{\mathrm{modes}}$ and $|T_{mn}|^2\approx0$ for $m\neq n$),
or at sufficiently small $L$, where the relative modal phases remain small, $e^{i(\beta_m-\beta_n)L}\approx1$.
For GRIN near $L=\ell L_{\mathrm{pitch}}$, all modes approximately rephase, increasing the shift-correlation range. This is confirmed by the calculated results in the main text (Fig.~2). In the main text, we also compared the width of $\langle M_E(\Delta\mathbf r,L)\rangle$ at $L=4.23~\mathrm{mm}$, where the modes are already strongly dephased in SI, with the standard ensemble-averaged field autocorrelation:
\begin{equation}
\begin{aligned}
\langle M_E^{\mathrm{auto}}(\Delta\mathbf r)\rangle
&=\langle \int E_{\mathrm{in}}^*(\mathbf r)\mathcal T_{\Delta\mathbf r}E_{\mathrm{in}}(\mathbf r)\,d^2\mathbf r\rangle
=\frac{1}{N_{\mathrm{modes}}}\sum_nT_{nn}(\Delta\mathbf r).
\end{aligned}
\label{eq:autocorrelation_weights}
\end{equation}

Its spatial scale is governed by average diagonal elements $T_{nn}(\Delta\mathbf r)$, which describe the overlap of a mode with its shifted copy and reflect the diffraction-limited spatial scale of the supported fields. The autocorrelation depends on linear weights $T_{nn}$, whereas Eq.~\eqref{eq:overlap_after_average} depends on squared magnitudes $|T_{mn}|^2$. These different weights help explain why the dephased translational correlation can be narrower than the field autocorrelation, even though both inherit spatial scales set by the supported modes.

\begin{figure*}[h]
\centering
 \includegraphics[width=1.00\textwidth]{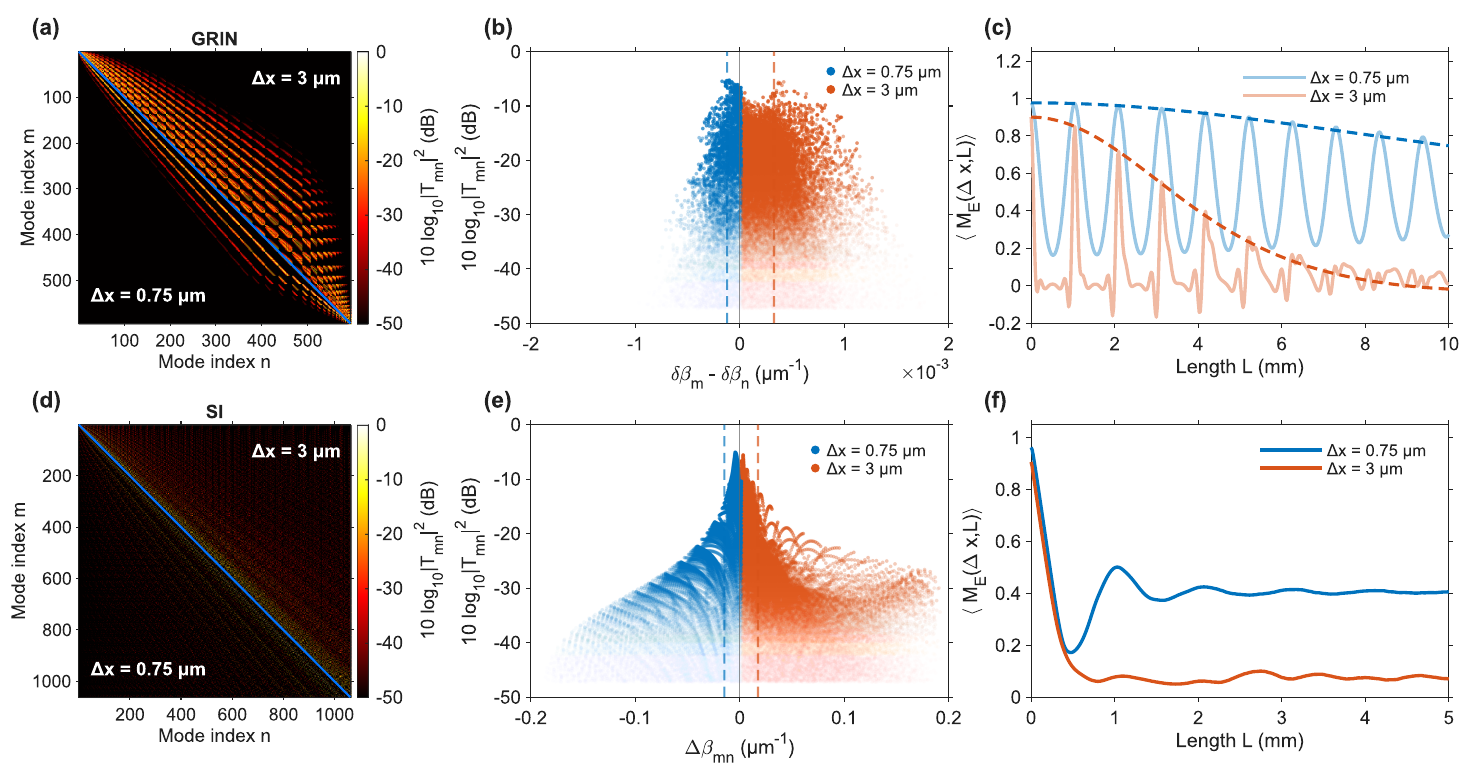}
\caption{Modal analysis of translational correlations in perturbed GRIN (a--c) and SI fibers (d--f) at two shifts $\Delta\mathbf r=(\Delta x,0)$: $\Delta x= 0.75~\upmu\mathrm m$ and $\Delta x= 3~\upmu\mathrm m$. (a,d) Translation-operator matrix elements $|T_{mn}|^2$ shown as heatmaps on a logarithmic scale. (b,e) The same elements plotted as a function of the corresponding propagation-constant differences $\Delta\beta_{mn}$ ($\delta\beta_{m}-\delta\beta_{n}$ for GRIN); dashed lines denote the weighted standard deviation of the distribution. Two shifts are shown in split view in (a,b,d,e) to simplify comparison. (c,f) Decay of $\langle M_E(\Delta\mathbf r,L)\rangle$ with $L$ (solid lines). Dashed lines in (c) correspond to the envelope, calculated by taking mismatches $\delta\beta_{j}$ instead of full propagation constants $\beta_{j}$ in Eq.~\eqref{eq:overlap_after_average}.}
\label{fig:memory}
\end{figure*}

Figure~\ref{fig:memory} shows an analysis of translational correlations. Typical distributions of $|T_{mn}|^2$ are shown as heatmaps in Fig.~\ref{fig:memory}(a,d). Because $|T_{mn}|^2$ is symmetric, we stitched matrices, corresponding to two shifts $\Delta x= 0.75~\upmu\mathrm m$ and $\Delta x= 3~\upmu\mathrm m$, along the diagonal to simplify comparison. For larger shifts, it is clear that more energy gets redistributed off-diagonal. The difference between GRIN and SI is also obvious: while the former has clear structure, the latter looks random. Nevertheless, there is a clear tendency of stronger intermode coupling near the diagonal. Since the modes, in this calculation, were indexed according to their propagation constants, this means that the highest contribution in $|T_{mn}|^2$ happens for smaller $\Delta\beta_{mn}$. To confirm it, we plotted the distribution of the matrix elements against the corresponding propagation constant difference (Fig.~\ref{fig:memory}(b,e)). For GRIN we considered the difference of propagation constant mismatches, since despite large accumulated phases the modes mostly rephase at the considered length $L=4 L_{\mathrm{pitch}}=4.23~\mathrm{mm}$. Clearly, the effective width of the distribution is much smaller than the maximal range that sets the decay scale of the input–output correlation $C(L)$. The translation-operator matrix elements $|T_{mn}|^2$ serve as weighting factors in the summation (Eq.~\eqref{eq:overlap_after_average}),
concentrating the weighted spectrum toward smaller propagation-constant differences $\Delta\beta_{mn}$
and extending the correlation length beyond $L_0\simeq2\pi/(\beta_{\max}-\beta_{\min})$. It is confirmed by the calculation of $\langle M_E(\Delta\mathbf r,L)\rangle$ as a function of $L$ in Fig.~\ref{fig:memory}(c,f). For SI, this increases the correlation length by more than a factor of ten. It is still, however, well under $1~\mathrm{mm}$, rendering this effect not applicable to a typical microendoscopy experiment. For GRIN, the effect is less prominent, but still noticeable ($\approx \times1.8$ at FWHM), keeping $\langle M_E(\Delta\mathbf r,L)\rangle$ substantial at $L=4.23~\mathrm{mm}$. This finding supports the use of translational correlations to improve the robustness of computational microendoscopy even when standard microscopy is no longer effective due to loss of input--output correlation.

\clearpage
\section{Compressive sensing simulation}
\label{sec:simulation_protocol}
Throughout this work, we estimated the mean performance of CS reconstruction and associated standard deviations for different metrics while varying two sources of error: measurement-matrix perturbation, determined by the lateral shift $\Delta\mathbf r$, and photon noise, determined by the total photon budget $N_{\mathrm{ph}}$. Successful CS recovery, however, depends on multiple conditions, including the number of measurements $M$. CS theory gives a characteristic scaling of the minimal required number of measurements $M\sim K\log(N/K)$ for an $N$-pixel image with $K$ significant coefficients in a sparse representation \cite{candesNearOptimalSignalRecovery2006a,donohoCompressedSensing2006}. The same scaling is justified by information-theoretic bounds in the presence of noise \cite{wainwright2009information}. For TV reconstruction, used in this work, reconstruction assumes sparsity of the image gradient, and corresponding recovery bounds are given in \cite{needell2013near}.

In our simulations, we estimated reconstruction quality by the Pearson correlation $Q=\rho(\vec x,\hat{\vec x})$ between the ground truth $\vec x$ and the reconstructed image $\hat{\vec x}$:
\begin{equation}
Q=\rho(\vec x,\hat{\vec x})=
\frac{\sum_{j=1}^N(x_j-\bar x)(\hat x_j-\overline{\hat x})}
{\sqrt{\sum_{j=1}^N(x_j-\bar x)^2\sum_{j=1}^N(\hat x_j-\overline{\hat x})^2}}.
\label{eq:pearson}
\end{equation}
When misalignment shifts $\Delta\mathbf r$ were considered, perturbing the measurement matrix, we used the shifted ground truth to study only the effect of decorrelation of speckle patterns without translation: $Q=\rho(\mathcal T_{-\Delta\mathbf r}\vec x,\hat{\vec x})$.

\begin{figure}[h]
\centering
 \includegraphics[width=0.97\linewidth]{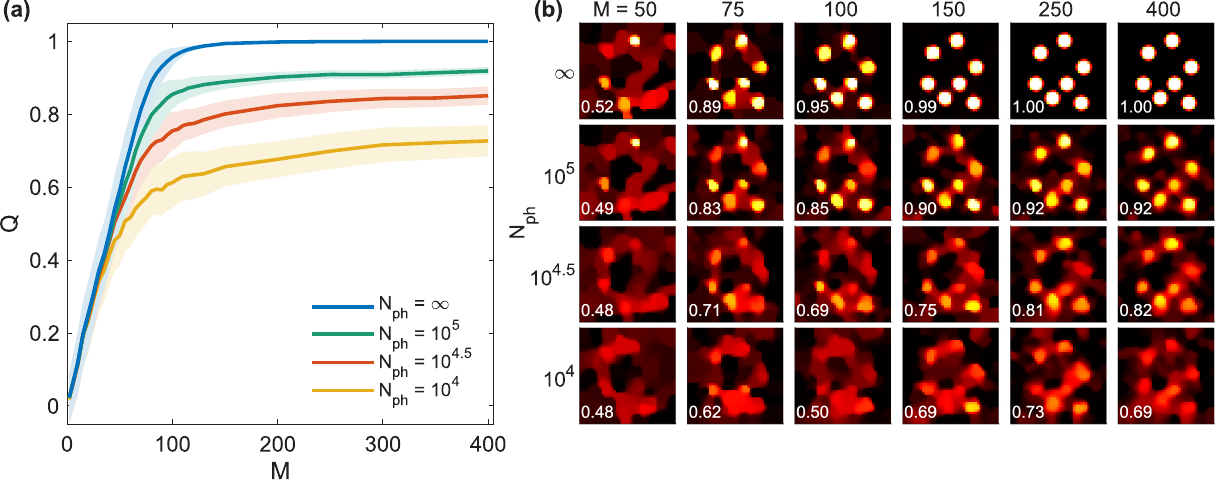}
\caption{Dependence of CS reconstruction on the number of illumination patterns in the presence of noise. (a) Mean Pearson correlation $Q$ versus $M$ for different expected detected photon counts $N_{\mathrm{ph}}$. Shaded bands indicate the standard deviation across realizations. (b) Representative reconstructed images. The corresponding $Q$ values are overlaid on the images.}
\label{fig:cs_bounds}
\end{figure}

The simulation started by fixing the field of view, number of pixels $N$ in it, and the object type. Then, we examined the dependence $Q(M)$ by adding illumination patterns and running $10$--$25$ trials. The mean and standard deviation are shown by the curves and semitransparent bands in Fig.~\ref{fig:cs_bounds}(a). Since the quality rises smoothly toward a plateau, we selected $M$ to reach $Q\simeq0.995$ for both fibers in the perfectly aligned, noise-free case. Once selected, $M$ was held fixed while the shift and photon budget were varied. This gives both fibers an almost perfectly reconstructed baseline from which to compare their sensitivity.

Additionally, we studied $Q(M)$ curves at different noise levels to confirm that one fixed value of $M$ can be sufficient for many simulations. The noise level is controlled by the total number of photons $N_{\mathrm{ph}}$ as explained below, with higher photon counts corresponding to lower relative noise. As can be seen from Fig.~\ref{fig:cs_bounds}(a), $N_{\mathrm{ph}}$ sets the maximum achievable $Q$, while the overall shape of $Q(M)$ stays mostly unchanged. Figure~\ref{fig:cs_bounds}(b) shows some reconstruction examples. In some situations, similar quality can be achieved by trading measurement count against photon budget (for example, $M=150$ at $N_{\mathrm{ph}}=10^5$ gives higher quality than $M=75$ at $N_{\mathrm{ph}}=\infty$ in Fig.~\ref{fig:cs_bounds}(b)), illustrating a tradeoff between the number of measurements and the detected photon budget, typical in many microscopy modalities. In CS, however, the rapid saturation of $Q(M)$ motivates choosing a sufficient measurement count near the plateau for all noise levels.

After fixing $M$, we can perturb the imaging model and study how it affects the reconstruction quality. The calibrated illumination patterns form the measurement matrix $A$, giving the ideal signal $\vec b=A\vec x$. To model an input shift, we propagate the shifted versions of the same input fields and obtain $A_{\Delta}=A+\delta A$. The corresponding expected signal is $\vec b_{\Delta}=A_{\Delta}\vec x$. To add photon noise at a chosen expected total detected photon count $N_{\mathrm{ph}}$, we scale $\vec b_{\Delta}$ so that the sum of its components equals $N_{\mathrm{ph}}$, generate independent Poisson-distributed counts, and then scale them back to the original signal units to get the measured signal $\vec b_{\mathrm{exp}}$:
\begin{equation}
\alpha=\frac{N_{\mathrm{ph}}}{\sum_i b_{\Delta,i}},\qquad
k_i\sim\operatorname{Poisson}(\alpha b_{\Delta,i}),\qquad
\vec b_{\mathrm{exp}}=\frac{\vec k}{\alpha},
\label{eq:photon_sampling}
\end{equation}
where $\vec k$ is the photon-count vector and $\vec b_{\mathrm{exp}}$ is the experimentally measured signal vector expressed in calibrated signal units. For the noiseless case, denoted $N_{\mathrm{ph}}=\infty$, we use $\vec b_{\mathrm{exp}}=\vec b_{\Delta}$ directly, since the relative Poisson error scales as $N_{\mathrm{ph}}^{-1/2}$.

To reconstruct the image, we use the original $A$ in the total-variation (TV) minimization problem of Eq.~(1) in the main text, and solve it using L1-magic \cite{L1Magic}. The mismatch seen by this reconstruction includes both the shift-induced change in signal and the photon noise, $\delta\vec b_{\mathrm{ph}}=\vec b_{\mathrm{exp}}-\vec b_{\Delta}$. In our simulations, we set the tolerance parameter $\epsilon$ to exactly match the residual associated with both errors:
\begin{equation}
\epsilon=\|\vec b_{\mathrm{exp}}-A\vec x\|_2
=\|\delta A\vec x+\delta\vec b_{\mathrm{ph}}\|_2.
\label{eq:combined_residual}
\end{equation}
In an experiment, this tolerance would instead have to be selected using a regularization criterion or an appropriate heuristic \cite{hansen1992analysis,santambrogioWhitenessBasedBilevelLearning2024,osherIterativeRegularizationMethod2005}.

The parameters used for Fig.~3 of the main text were chosen as a tradeoff between computation time and relevance to practical microendoscopy. The $16\times16~\upmu\mathrm{m}^2$ field of view fit entirely within the fiber core and was sampled on a $40\times40$ grid ($N=1600$), with a pixel size of $0.4~\upmu\mathrm{m}$. We used $M=80$ measurements, corresponding to $5\%$ of the reconstruction pixels. The Nyquist sampling pitch $\lambda/(4\mathrm{NA})\approx0.60~\upmu\mathrm{m}$ dictates approximately $700$ sampling points across this field of view. Thus, the measurement number $M$ was approximately $11\%$ of this Nyquist estimate. The diffraction-limited resolution was approximately $1.5~\upmu\mathrm{m}$. Each simulated object contained three beads with a diameter of $2.4~\upmu\mathrm{m}$, with approximately $8.4\%$ of the image pixels having nonzero values. For each shift and photon budget, reconstruction quality was averaged over $25$--$150$ trials, depending on the noise level.

\clearpage
\section{Reconstruction examples and other quality metrics}
\label{sec:negative}

\begin{figure}[!b]
\centering
 \includegraphics[width=0.97\linewidth]{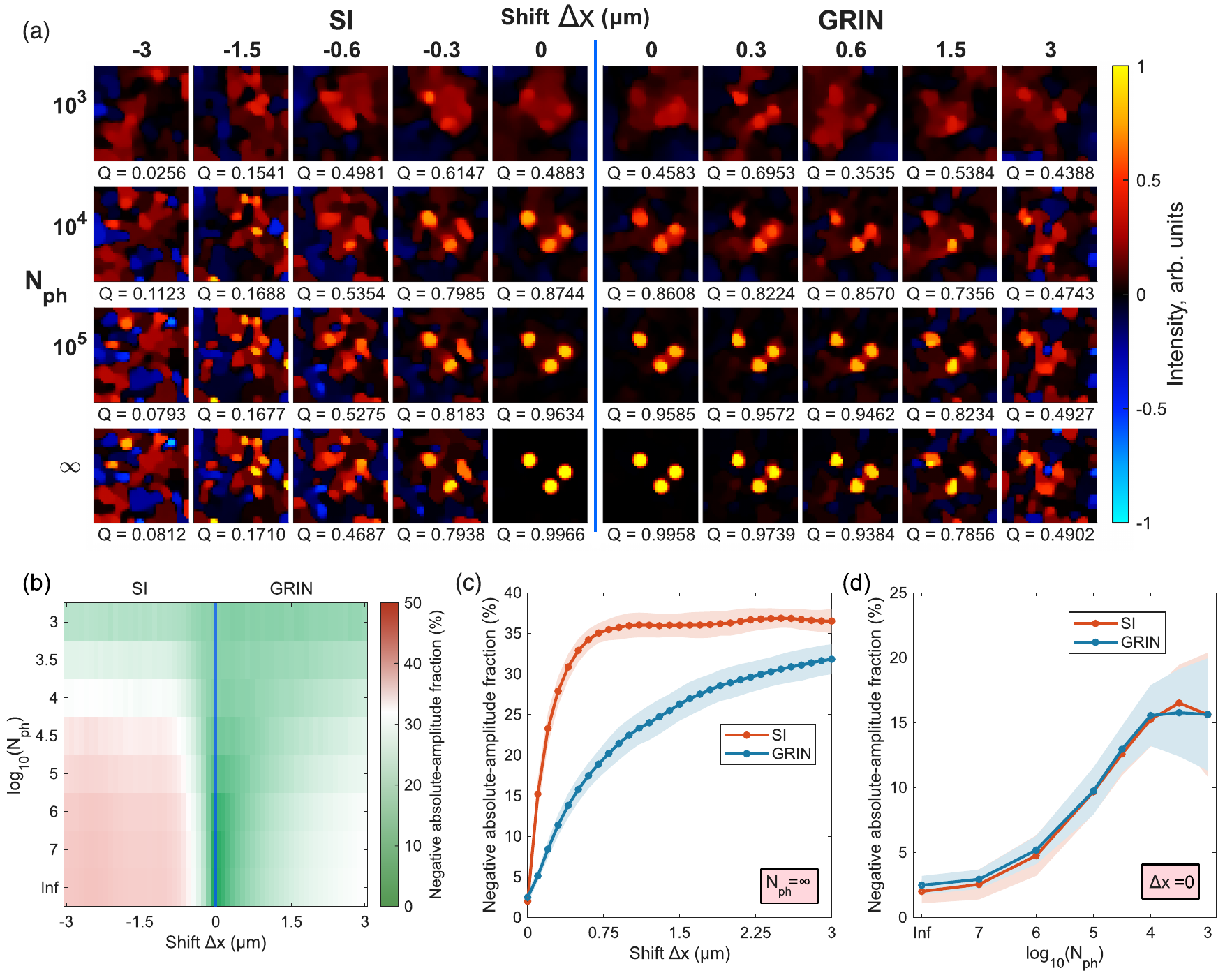}
\caption{Reconstruction examples and the negative-intensity fraction under different photon budgets and lateral shifts. (a) Representative reconstructions comparing SI (left) and GRIN (right). The shifts and photon budgets $N_{\mathrm{ph}}$ are indicated by the columns and rows, respectively. The Pearson correlation of the reconstructed image with the ground truth is indicated below each image. (b) Heatmap and its cross-sections at zero noise (c) and zero shift (d) of the intensity-weighted negative fraction $f_-$ in percent.}
\label{fig:negative}
\end{figure}

Figure~\ref{fig:negative}(a) shows reconstruction examples corresponding to Fig.~3 of the main text. The grid of images demonstrates how the reconstruction quality decreases with larger shifts and lower photon budgets. The GRIN fiber retains recognizable details even at a shift of $3~\upmu\mathrm m$, whereas the SI reconstructions deteriorate at substantially smaller shifts.

In an experiment, the ground truth is not known and Pearson correlation $Q$ cannot be maximized to compensate for an unknown shift. Another metric must be used to distinguish noise artifacts from signal based on the reconstructed image alone. Although fluorescence intensity is nonnegative, the TV reconstruction can assign negative values to some pixels. For a nonnegative image, the summed pixel intensity is proportional to the total reconstructed optical signal. Negative pixels represent an unphysical negative contribution to this signal. We quantify its relative weight, for a nonzero reconstruction, as an intensity-weighted fraction:
\begin{equation}
f_-={\displaystyle\sum_{j:\,\hat x_j<0}|\hat x_j|}\ /\
{\displaystyle\sum_j|\hat x_j|}\times100\%.
\label{eq:negative_fraction}
\end{equation}
The numerator is the sum of absolute values of negative pixels and the denominator is the total magnitude of all reconstructed pixels.

Figure~\ref{fig:negative}(b) presents the heatmap of $f_-$ as well as its cross-sections (Fig.~\ref{fig:negative}(c,d)) as a function of lateral shift and photon budget, analogously to Fig.~3 of the main text. This metric is minimal near zero shift and at high photon budgets, where the reconstruction quality is close to perfect (Fig.~\ref{fig:negative}(b)). Over a broad parameter range, GRIN gives a lower $f_-$ than SI. This dependence suggests that minimizing $f_-$ could help guide a manual or automatic iterative shift-compensation process.
While a dependence on shift remains visible in GRIN up to $3~\upmu\mathrm m$, the SI curve reaches a plateau already at $\approx0.75~\upmu\mathrm m$ (Fig.~\ref{fig:negative}(c)), leaving a smaller shift range over which this metric could guide alignment.
The shot-noise dependence of $f_-$ is practically indistinguishable for both fibers (Fig.~\ref{fig:negative}(d)), in accordance with the behavior of $Q$ (Fig.~3(c) of the main text).

\footnotesize
\bibliographystyle{opticajnl}
\bibliography{paper_2_curated_upd3}